\documentclass{article}
\usepackage[T1]{fontenc}
\usepackage{lmodern}
\usepackage{graphicx}
\usepackage{epsfig}
\usepackage{bbm}

\begin{document}

\title{Gauge Symmetry, Spontaneous Breaking of Gauge Symmetry: Philosophical Approach }

\author{P. Lederer, \\
Directeur de recherche \'em\'erite au CNRS\\
LPS, Bat. 510, Universit\'e Paris-Sud, 91405-Orsay cedex\\
Tel: 33662984051 --- pascal.lederer@u-psud.fr}

\maketitle
\begin{abstract}
Abstract

This paper deals with the Berry phase, and the ontology of the electromagnetic  vector potential. When the state of the system is gauge symmetric, the  vector potential may be interpreted as a convenient tool of a mathematical formulation, with no ontological meaning.
 I argue that this interpretation  is in difficulty because the vector potential depends linearly on  the supercurrent in the superfluid state, which is a spontaneously broken gauge symmetry state, where particle number is not conserved. I suggest that when gauge symmetry is spontaneously broken, the vector potential becomes an emergent material object of nature.
The revised version includes  sections on scientific realism, and emergence, and  new references on Noether's theorem, among others.
\end{abstract}

\vspace{20mm}
\section{Introduction}

This paper addresses the question of the ontology of the vector potential. This  question   has been in debate among physicists ever since Maxwell introduced  the vector potential  to account for the Faraday effect. The magnetic field is the curl of the vector potential. The latter may be shifted by adding to it   a vector field the curl of which vanishes. Thus infinitely many vectors  correspond to  the vector potential, while the magnetic field is unique. This is what is called gauge symmetry. Theories which are formulated in terms of potentials are ``gauge theories''.  These acquire growing importance in theoretical physics, in particular in condensed matter, elementary particle physics and cosmology. 

Spontaneous symmetry breaking is also a  common feature of many fields of physics: the ground state of a many-particle system may be invariant  under the operations of a subgroup of the total symmetry group of the Hamiltonian. This is generally linked to the occurrence of a phase transition, for example when the temperature increases, from an ordered state, with low symmetry, to a less ordered state with higher symmetry. For example, a ferromagnetic order has axial symmetry around the magnetization vector. A sufficient  increase in temperature triggers a transition, at a critical temperature $T_c$  to a paramagnetic state which has the full rotation symmetry, which is the symmetry group of the Hamiltonian.
  
 I would like to connect those two topics and discuss some of the lessons we can learn about the material world,  and knowledge of its laws,  by examining various aspects of gauge theories. The latter  are relevant in classical physics, and  are connected in quantum mechanics  through the phase of the wave function.  The ontological status of the phase of the wave function may be clarified by discussing some concepts such as the Berry phase  and    particular cases of spontaneous gauge symmetry breaking, i.e. the phenomena of superfluidity and superconductivity. The relationship of gauge symmetries  with electricity conservation is especially clear in the latter phenomena, where the breaking of gauge symmetry is associated with the canonical conjugation of phase and particle number.  

Section \ref{classic} below will review some basic notions on gauge symmetry in classical physics. Section \ref{qm} describes the quantum mechanical version of this topic,  discusses the Aharonov Bohm effect in section \ref{ab} and the Berry phase  in section \ref{berry}. Both are relevant experimental and theoretical topics  for my purpose. Section \ref{supercond} discusses some aspects of superconductivity ,  a phase with spontaneously broken gauge  symmetry

The discussion among philosophers about the topics mentionned above revolves around central questions of knowledge: how can the human mind  access truths about the world? Are theories in physics mere representations of phenomena? Are they able to reflect, in a more or less exact way, real processes of the world? Are theoretical entities such as fields, particles, potentials, etc., real objects, independent of the human mind? This will be discussed in section \ref{sr}. Should eventually all theories of physics reduce to one fundamental Theory of Everything? Or are there emergent properties which have qualitative features absent from the fundamental microscopic  equations? Emergence will be discussed in section \ref{emerg}. 

Proposals and ideas expressed in this paper are summarized in section \ref{conclu}.

\section{Classical physics} \label{classic}
   Two topics are of interest in this chapter, that of Maxwell equations for classical electrodynamics, and that of parallel transport, such as is at work in the Foucault pendulum
\subsection{Maxwell equations, the Faraday effect, and the vector potential}
The electric field $\vec{E}(x,t)$ and magnetic field $\vec{B}(x,t)$ obey the four Maxwell equations, which describe  how the fields are related to one another and to static or moving charges.The four Maxwell equations are:
\begin{eqnarray}\label{maxw}
(a)~~~~div\vec{E} =\rho~~~~~~~~~ &~~~~~~~~~~~~~~&(c)~~~~div\vec{B}=0\\
(b)~~~~curl\vec{B}-\frac{\partial \vec{E}}{\partial t}=\vec{j}&~~~~~~~~&(d)~~~~curl \vec{E}+\frac{\partial \vec{B}}{\partial t}=0
\end{eqnarray}
where $\rho$ is the charge density and $\vec{j}$ is the current density.

  Quantities such as the electric potential $V(x,t)$ and the vector potential $\vec{A}$ are usually labeled "auxiliary quantities". They determine completely $\vec{E}$ and $\vec{B}$ according to:
\begin{equation}\label{gauge1}
\vec{B}= \vec{\nabla} \wedge \vec{A} \hspace{2cm}  \vec{E} = - \frac{\partial}{\partial t}\vec{A} -\vec{\nabla}V
\end{equation}
On the other hand, $\vec{B}$ and $\vec{E}$ do \underline{not} determine $V$ and $\vec{A}$.  If $f$ is a scalar function of space and time, the following transformation on the potentials, a "gauge transformation",  does not alter the fields\footnote{This is due to equations (\ref{gauge1}) and to $\vec{\nabla} \wedge \vec{\nabla} =0$.}:
\begin{equation}\label{gauge2}
\vec{A} \rightarrow \vec{A} + \vec{\nabla} f  \hspace{2cm}  V \rightarrow V -\frac{\partial}{\partial t}f.
\end{equation}
This is the essence of gauge invariance, or gauge symmetry. 
Is this merely an ambiguity of the mathematical representation of physical states?  A mere representation surplus? References \cite{guay,healey} are  examples of philosophical investigation of gauge theories. 

 David Gross \cite{gross} comments on the way Maxwell introduced the vector potential in order to account for the Faraday effect. The latter is the occurrence  of an electric current  in a closed conducting loop when the magnetic flux threading the loop varies in time.
Maxwell did not accept the non locality of the effect: consider a situation such that the magnetic field is concentrated in a thin cylinder at the center of a closed conducting loop, and vanishes elsewhere.  How could an electric current be induced by a magnetic flux variation far away from the loop, with  zero intensity of the field at its locus? Maxwell found a satisfactory solution by inventing the vector potential $\vec{A}$. The latter has non zero values in regions where the magnetic field vanishes. The time variation of the flux through the loop could now be ascribed  to $\partial_t\vec{A}$, together with the relationship of the electric field with the time variation of $\vec{A}$: $\vec{E}= -\partial_t \vec{A}$. The electric field then acts locally on the metallic loop, where the vector potential is non zero: a local description of phenomena is retrieved.

   There was no doubt in Maxwell's mind that $\vec{A}$ was a physical (i.e. real)  field.
	
	But a problem appears with gauge invariance, as exhibited by equation (\ref{gauge2}). How can a  physical object exist if it can be described by  an infinite number of different vector fields? What led Maxwell to think of the vector potential as a physical field was the actual non zero value of $\frac{\partial}{\partial t} \vec{A}$ at the locus of the conductor, and the gauge invariance of the circulation of $\vec{A}$  along a closed loop C. This is an example of holonomy, i.e. a property of a closed loop. Indeed,$\oint_C \vec{A}.\vec{ds}$ is, through Stokes theorem, equal to the flux  $\Phi_C$ of $\vec{B}$ through the loop C.  It is indeed gauge invariant since the circulation of a gradient on a closed loop is identically zero:  $\oint_C \vec{\nabla f}.\vec{ds}= \int_S \vec{\nabla}\wedge \vec{\nabla}f dS=0$ ($S$ is the surface subtended by the closed curve $C$). 
	
	In the presence of charges and currents, Maxwell's equations impose the conservation of charge. In the classical theory of electromagnetism, the connection between gauge invariance and charge conservation was only realized in 1918 by Emmy Noether's first theorem \cite{noether}, as well as Weyl's attempts to construct a unified  theory of gravitation and electromagnetism\cite{Weyl}. This connection will be discussed in another section of this paper. 
	
	The positivist attitude towards science ( as  clearly expressed, for example, by Duhem \cite{duhem} or Mach \cite{mach}  ) prevailed among many of Maxwell's followers: Hertz, Heaviside, Lorentz, etc.,  down to Aharonov \cite{Aharonov}. This posture is that of Cardinal Bellarmin, who approved of Galileo's as long as the heliocentric hypothesis allowed to account for phenomena\cite{duhem}\footnote{Duhem strongly advocates Bellarmin's position , and would have condemned Galileo...} ({\it{"sauver les apparences"}}), but prohibited drawing ontological inferences about the world. Many physicists have adopted the view that the vector potential is a practical tool to simplify Maxwell's equations and to account for phenomena, but has no physical meaning, no ontological content. The rationale behind this view is the gauge dependent nature of $\vec{A}$ which makes it unobservable\footnote{Observable, or unobservable, is used here in the physicists' way: an observable thing which has causal powers allowing it to produce detectable effects, whether by a signal on a screen, a trace on a chart, or by other technologies available to the experimentalist; can an unobservable thing (in the physicist's meaning of the word) be real? Physicists in general dismiss this question as meaningless...} as a local quantity. As we shall see, a spontaneous gauge symmetry breaking seems to turn this "unobservable" object into a directly measurable one.
	
	Gauge theories have been recently discussed by Healey \cite{healey} from a philosophical viewpoint. Healey has concentrated on the Non Abelian Yang Mills theories which appear in the standard model. As this discussion involves specialized notions, such as "`soldering forms of fiber bundles"', which turn the reading of his book rather arduous,  I will not discuss Healey's work in detail. I only summarize his position, namely that Yang-Mills theories refer to nonlocal properties encoded in holonomies,  and the local gauge symmetries that characterize them are purely formal and have no direct empirical consequences. Brading and Brown \cite{bb} on the other hand insist  that  Noether's first theorem, which will be discussed  below establishes that the {\it{very fact that a global gauge transformation does not lead to empirically distinct predictions is in itself  non trivial.}} They state that {\it{the freedom in our descriptions is no 'mere' mathematical freedom -- it is a consequence of a physically significant structural feature of the theory.}} A rather easy introduction to gauge invariance can be found in a Field Theory treatise such as  reference \cite{ramond}.
	
	As will be clear in the following, my own attempt at discussing the simpler Abelian gauge theories within non relativistic quantum mechanics deals with the same topics: are there conditions for which gauge potentials can be reasonably associated with a real material object?

	The philosophical background of this paper might be called the question of scientific realism: are the theoretical  entities which appear in the course of science real? Under what conditions can we have good reasons to believe in the reality of a theoretical entity? Are electrical or magnetic fields real? Are potentials real? Are physical laws described by theories real? A related question is that of materialism: following the latter, the human mind, and the mind independent reality are both different forms of matter; can the human mind, in its individual or social form, access in principle knowledge of true aspects of matter?
	
	Those questions are discussed at more length in section \ref{sr}.
	
	The classical Hamilton function $H$ for a single charged particle in the presence of potentials is expressed as:
	\begin{equation}\label{hamilton}
	H=\frac{1}{2m}\left( \vec{p} -e\vec{A}  \right)^2  +eV
	\end{equation}
	where $\vec{p}$ is the canonical momentum.
	
	The dynamic equation describing the motion of a charged classical particle is the Lorentz equation, which can be derived from equation (\ref{hamilton}):
	\begin{equation}\label{lorentz}
	m\frac{d^2 \vec{r}}{dt^2} = e\vec{E} + e\left( \frac{d\vec{r}}{dt} \right)\wedge \vec{B}
	\end{equation}
	The hamiltonian  (equation (\ref{hamilton})) is expressed in terms of the potentials, while equation (\ref{lorentz}) is expressed in terms of the fields. The choice, in classical physics, is a matter of taste.
	
\subsection{Parallel transport, and the Foucault pendulum}\label{parallel}

The discovery of  the quantum mechanical Berry phase \cite{berry}, discussed in section \ref{qm}  in this paper,  has allowed to re-discover a hitherto little studied gauge invariance connected, in classical physics, to parallel transport.

Two concepts are of interest in this topic: the concept of {\it{anholonomy}} and that of  {\it{adiabaticity}}. Quoting Berry \cite{berry}; {\it{" Anholonomy is a geometrical phenomenon in which nonintegrability causes some variables to fail to return to their original values when others, which drive them, are altered through a cycle...Adiabaticity is slow change and denotes phenomena at the borderline between dynamics and statics"}}. I note in passing that adiabaticity is yet another concept which supersedes the traditional text book  antinomy between statics and dynamics.

The simplest example of anholonomy is the change of the direction of the swing of a Foucault pendulum after one rotation of the earth. Visitors to the Panth\'eon in Paris can check that this is a  phenomenon at work in the objective real world.

If a unit vector $\vec{e}$ is transported in a parallel fashion over the surface of a sphere, its direction is changed  by an angle $\alpha (C)$ after a closed circuit $C$ on the sphere has been completed. 
 $\alpha (C)$ is found to be  the solid angle subtended by $C$ at the center of the sphere. It is expressed by the circulation of a certain vector on $C$, the result of which is independent of the choice of basis vectors\footnote{For details of the derivation see ref. \cite{berry}.}. The latter   freedom of choice is a gauge symmetry, the change of basis vectors being equivalent to a change of gauge. This feature is analogous to what we will find in quantum mechanics, either when  discussing adiabatic transport of a quantum state, or the electromagnetic vector potential: an objective phenomenon of nature depends on the circulation of a vector quantity along a closed loop, although  that quantity, when gauge invariance prevails, cannot be defined at any  point along the circuit.

Before going over to quantum mechanics, let me discuss first some ideas which are implicit in what I have written above, namely my scientific realist point of view.

\section{Scientific Realism and Materialism}\label{sr}

In his interesting  book {\it{Representing and Intervening}}\cite{hacking}, Ian Hacking critically reviews a number of positivist or agnostic philosophers, from Comte to Duhem \cite{duhem}, Kuhn\cite{kuhn}, Feyerabend\cite{feyer}, Lakatos\cite{lakatos}, van Fraassen\cite{frassen}, Goodman\cite{goodman}, Carnap,    etc.. I define positivism here, loosely, as the   philosophical thesis  which reduces knowledge to establishing a correspondence  between theories, or mathematical symbols,  and phenomena, and denies that it may access ontological truths, dubbed "`metaphysics"'. Hacking  writes (p.131), about various trends of positivism: {\it{ Incommensurability, transcendental nominalism, surrogates for truth, and styles of reasoning are the jargon of philosophers. They arise from contemplating the connection between theory and the world. All lead to an idealist cul-de-sac. None invites a healthy sense of reality...By attending only to knowledge as representation of nature, we wonder how we can  ever escape from representations and hook-up with the world. That way lies an idealism of which Berkeley\footnote{Berkeley was the finest example of philosophical idealism, which in his case is solipsism.  Kant rejected it  in favor of transcendental idealism.} is the spokesman. In our century (the twentieth) John Dewey has spoken sardonically of a spectator theory of knowledge...I agree with Dewey. I follow him in rejecting the false dichotomy between acting and thinking from which such idealism arises...Yet I do not think that the idea of knowledge as representation of the world is in itself the source of that evil. The harm comes from a single-minded obsession with representation and thinking and theory, at the expense of intervention and action and experiment}}.

I  agree with Hacking, inasmuch as he defends scientific realism. Scientific realism says that the entities, states and processes described by correct theories really do exist.  I do not underestimate, as I think Hacking seems to do, the explanatory power of a correct theory, such as Maxwell's theory for electromagnetism. But  I believe that Hacking is fundamentally correct in stating that the criterion of reality is practice. He describes a technique which uses an electron beam  for  specific technical results. This convinces him that electrons exist. He thinks that "`{\it{reality has more to do with what we do in the world than with what we think about it.}}"`He discusses at length experiments, and points out that experimenting is much more than observing: it is acting on the world, it is a practical activity. The certainty I have about the reality of  a magnetic field originates from the experiments, observations, theory\footnote{Observation here is literally seeing a certain intensity value  on a screen or on a chart connected by electric leads to a conducting solenoid; or observing how a charged particle motion is deflected when a magnetic field is turned on, etc..}, and practice\footnote{Practice ranges from using a compass for sea travel to Nuclear Magnetic Resonance used in medical imagery, for example.}. This certainty is intimately connected, not only with Maxwell's equations, but also with various historical acquisitions of physics, mostly during the nineteenth century; for example the certainty that those equations describe correctly a vast amount of electromagnetic phenomena, which are at the basis of countless technologies which billions of humans use everyday,  which govern a  vast amount of industrial production, etc.. 

Hacking points out that not all experiments are loaded with theory, contrary to statements by Lakatos \cite{lakatos}. Maxwell's equations belong to this category of discoveries  where experiments were intimately intertwined with theory. 

 Consider the following quotation: "`{\it{The question whether objective truth can be attributed to human thinking is not a question of theory but it is a practical question. Man must prove the truth --i.e. the reality and power, the this-sidedness of his thinking in practice. The dispute over the reality or non reality of thinking that is isolated from practice is a purely scholastic question}}"'. That is Marx' thesis 2 on Feuerbach \cite{marxfeuer} published in 1845! A second quotation is also relevant: "` {\it{The result of our action demonstrates the conformity ($\ddot{U}$bereinstimmung) of our perceptions  with the objective nature of the objects perceived}"'}. That is due, in 1880,  to Engels \cite{engelspractice} who is also the author of a well known expression: "`{\it{The proof  of (the reality of)  the pudding is that you eat it}} "`. Compare with Hacking (p.146 of \cite{hacking}): "`{\it{"`Real"' is a concept we get from what we, as infants, could put in our mouth}}"'  

How come Hacking does not refer to those predecessors who have stressed, as he does, the practice criterion as the criterion of reality?

The answer is probably in p.24 of Hacking's book quoted above \cite{hacking}: "`{\it{...realism has, historically, been mixed up with materialism, which, in one version, says everything that exists is built up out of tiny material blocks...The dialectical materialism of some orthodox Marxists gave many theoretical entities a very hard time. Lyssenko rejected Mendelian genetics because he doubted the reality of postulated genes}}"'.

It is a pity that Hacking, who reviews in many details all sorts of nuances between various doctrines he eventually calls idealist, dismisses materialism on the basis of a version of materialism ("`of some orthodox Marxists"') long outdated. As for his dismissal of dialectic materialism, he is certainly right in condemning its dogmatic degeneracy during the Stalin era. But is this the end of the story?

Dialectical materialism -- or at least its reputation --  suffered a severe blow, as a useful and rational philosophical system, when it was used as official state philosophy in the USSR.  Much to  the contrary, nothing,  in the founding philosophical writings \cite{marx,engels,lenin} allowed to justify turning them into an official State philosophy. One may surmise that the striking political achievements of the first years of the Soviet revolution made that appear as a positive  step. This produced however such catastrophies as  the State support for Lyssenko's theories, based on the notion that genetics was a bourgeois science, while lamarckian concepts were defined at the government level as correct from the point of view of a caricature of dialectical materialism. It is understandable that such nonsense in the name of a philosophical thesis turned the latter into a very questionable construction in the eyes of many.

Dialectic materialism itself is an open system, which has no lesson to teach beforehand about specific objects of knowledge, and insists \cite{engels} on taking into account all lessons taught by the advancement of science.

 It is perhaps time for a serious critical assessment of this philosophical thesis. The question of the possibility of general theoretical statements  about the empirical world is not a negligible one. 

 Anyone who would dismiss Hacking's positions on realism,  on the basis that he made an erroneous statement about quantum mechanics \footnote{P. 25 of \cite{hacking}, one reads: "`{\it{Should we be realists about quantum mechanics? Should we realistically say that particles do have a definite although unknowable position and momentum?}}"'. In fact the very classical concept of trajectory, with simultaneously well defined  position and momentum is invalid for a microscopic particle.  Particles do not have simultaneously definite position and momentum. In a propagation process, a particle follows simultaneously different trajectories, which is an aspect of the superposition principle. Realism about quantum mechanics is justified, and natural as soon as one admits that classical behaviour is, in general, an approximation valid for actions large compared to $\hbar$. } would certainly not do  justice to his philosophical views. 

Hacking distinguishes between realism about theory and realism about "`entities"' (atoms, electrons, quarks, etc.), e.g. p. 26 of reference \cite{hacking}:"`{\it{The question about theories is whether they are true or are true-or-false...The question about entities is whether they exist }}"'. He is a realist about "`entities"' but doubts realism about theories. In my view, electromagnetism is a good example where realism about theory and realism about "`entities"' (magnetic or electric  fields, currents and charges, etc.) are both relevant. Take one example of well known technology:  magnetic fields,  nuclear spins in animal tissues,  interactions of the latter with microwave radiations, manage  to provide images of the interior of our body. The latter allow sufficient accuracy that tumors, for example, having thus been made "`observable"' on photographs or  fluorescent screens,  can be efficiently removed by surgery. Does this allow doubts to persist about the reality of magnetic fields, nuclear spins, or electromagnetic radiation? Theories explaining the behaviour of nuclear spins, their interaction with magnetic fields and with electromagnetic radiation, and eventually with fluorescent screens, etc., must have, with a certain degree of accuracy, within certain ranges of experimental parameters,   an undisputable truth content. 

However  Maxwell's theory of electromagnetism might also be said to be false, since it ignores the quantum mechanical aspects of light. Can it be both true and false? Is the distinction between realism about theories and realism about "`entities"' a fully rational one?

Dialectical materialism offers an interesting view. First, contrary to the caricature  mentioned above, materialism gives a clear answer to the "`{\it{gnoseological problem of the relationship between thought and existence, between sense-data and the world...Matter is that which, acting on our senses, produces sensations.}} This was written in 1908 by Lenin \cite{lenin}, far from Hacking's  caricature quoted above. It may look too simple when  technology  is intercalated between matter  and the screens on which we read experimental results. As discussed by Bachelard \cite{bachelard} and Hacking, a reliable laboratory apparatus is a phenomenon operator which transforms causal chains originating from the sample into readable signals on a chart or on a screen. Technology or not, matter is the external source of our sensations. So much for materialism. Dialectical materialism adds  a fundamental aspect of matter i.e. that contraries coexist and compete with each other within things in Nature. Depending on which dominates the competition (contradiction) under what conditions, the causal chains originating from the thing and causing phenomena will take different forms, which are reflected in theories. Epistemics and ontology are intimately intertwined. This is why Maxwell's theory of light is both true and false. In all electromagnetic  wave like phenomena it has a definite truth content.

  Theories undergo a complex historical process of improving, sometimes correcting qualitatively, representations  of how the things are. The theory of electromagnetism from Maxwell's treatise \cite{maxwell} from 1873 to this day is a good example. Maxwell thought, and many a physicist of his time with him, that his treatise meant the final point for physics. Quantum mechanics, special relativity, general relativity, condensed matter physics, atomic physics, astrophysics, high energy physics suggest on the contrary that the progress of knowledge of nature is inexhaustible. This is Chalmers' point of view \cite{chalmers}, for example. It is tightly associated with technological improvements which are themselves the results of scientific advances.
	
	Some theories may prove false. Somme theories may be true.  But most  good theories are not, in general,  either true or false. Parameters and orders of magnitude have to be specified.

\section{Quantum Mechanics} \label{qm}

\subsection{Electromagnetic gauge symmetry}

The ontological question about the vector potential was revived when it became clear that the Schr\"{o}dinger equation for charged particles in the presence of a magnetic field had to be formulated in terms of the  potentials $\vec{A}$ and $V$ (apart from the Zeeman term which will be dropped here from the picture for simplicity, with no loss of generality \footnote{This means that we are  interested here with orbital degrees of freedom, as if spin degrees of freedom  were frozen in a large enough field}). The reason is that  there is no such choice as between equations (\ref{hamilton}) and (\ref{lorentz}). The only starting point in quantum mechanics  is the Hamiltonian\footnote{as opposed to the classical case when one could start with the Lorentz force equation. The lagrangian formulation also deals only with potentials; the choice between the lagrangian formulation or the hamiltonian one is a technical matter. } which is given by (\ref{hamilton}) where $\hat{\vec{p}}$ is now an operator: $\hat{\vec{p}} = -i\hbar \vec{\nabla}$.
Quantum mechanics substitutes the notion of quantum state to that of the classical notion of trajectory.  The latter  is irrelevant at the microscopic level, as evidenced by the non commutativity of momentum $p$ and coordinate $x$.

Gauge invariance is now expressed by the simultaneous transformation of equation (\ref{gauge2}) with:
\begin{equation}\label{gauge3}
\Psi \rightarrow \Psi \exp{i f(x, y, z, t)},
\end{equation}
where$\Psi$ is a solution of the time dependent  Schr\"{o}dinger equation. Thus the state is described up to a phase factor.

Most quantum mechanics text books, at least up to Berry's discovery in 1984, state that the overall phase factor of the wave function describing a system has no physical meaning, since $|\Psi|^2$ is unchanged when the overall phase changes\footnote{see for example ref.\cite{landau}.}. Since $|\Psi(\vec{r})|^2$ is the particle density at site $\vec{r}$, the density, as well as the total particle number are invariant under a phase change. This is a quantum version of the charge conservation described by Maxwell's equations: a global phase change, which is a global gauge change, conserves the charge.  We shall see the consequences of that statement: what if a quantum state breaks gauge invariance?

 At first Weyl \cite{Weyl} linked charge conservation to local gauge transformations. The latter are "local" when the gauge shift $f(\vec{r},t)$ varies in space. In fact Noether \cite{noether} showed that global gauge invariance is enough to express charge conservation\footnote{see ref. \cite{brading} for a detailed discussion.}. Global gauge is the limit of a local gauge when $f$ in equation \ref{gauge3} is a constant. Gauge symmetry is thus seen to have a profound significance and cannot be reduced to a 'mere' representation surplus: it is a fundamental symmetry of the material world.

As regards the "representation surplus" aspect of gauge freedom, it is worth pointing out that this surplus is a blessing for the theorist, because it allows a mathematical treatment of problems which is adapted to the specific geometrical features at hand. The behaviour of the electronic liquid under magnetic field in a long flat ribbon is conveniently expressed in a gauge where $\vec{A}$ is orthogonal to  the long dimension of the ribbon. For the physics of a disk, a gauge with  rotational symmetry is usually useful. Any gauge choice should yield the same result, but a clumsy choice can make the theory intractable. 
This is quite analogous to the correct choice of coordinate system -- cartesian, polar, cylindrical, spherical, etc. -- in a geometry problem. At this stage, considering  the vector potential as a mere technical tool -- a usefully flexible one at that -- for the theory seems  rational.

\subsection{The Aharonov-Bohm effect} \label{ab}

The Aharonov-Bohm effect \cite{AB} proves that there exist effects of static potentials on microscopic charged particles, even in the region where all fields vanish. The standard experimental set up may be described by the diffraction of electrons by a standard two slits display. An infinitely long solenoid,  is placed  between the two slits, parallel to them, immediately behind the slit screen; an electric current creates a magnetic flux inside the solenoid, and none outside. The electronic wave function is non zero in regions where no magnetic flux is present. A variation of the flux inside the solenoid causes a displacement of the interference fringes on a second screen placed behind the slits.  
It is straightforward to relate this displacement to the phase difference $\delta \phi$ of the two electron paths at a given point on the screen. The latter is given by
\begin{equation}
\delta \phi \propto \frac{e}{\hbar}\oint_C \vec{A}.\vec{ds} = e\Phi_B/\hbar \equiv  2\pi \Phi_B/(h/e)
\end{equation}
where $\Phi_B$ is the flux threading the solenoid. This flux is gauge invariant. The displacement is periodic when the flux $\Phi_B$ varies in  the solenoid, with a period given by the flux quantum\footnote{In this paper the velocity of light, c,  is put equal to unity throughout.} $h/e$.

There are various other versions of the same effect. One is the observation of periodic variations -- with flux period $2\pi h/e$ --  of the resistance of a mesoscopic conducting ring when the flux varies inside a thin solenoid passing through the ring \cite{Aharonov}. Yet another variant will be discussed  in a later section  when I discuss superconductivity.
In their 1959 paper \cite{AB}, Aharonov and Bohm discuss the ontological significance of their findings. They mention that potentials have been regarded as purely mathematical objects. Quoting them: {\it{... it would seem natural to propose that, in quantum mechanics, the fundamental natural entities are the potentials, while the fields are defined from them by differentiation. The main objection ...is grounded in the gauge invariance of the theory...As a result the same physical behaviour  is obtained from any two potentials $\vec{A}$ or $\vec{A'}$ (related by a gauge transformation). This means that insofar as the potentials are richer in properties than the fields, there is no way to reveal this actual richness. It was therefore concluded that the potentials cannot have any meaning, except insofar as they are used mathematically, to calculate the fields.  }} Over the years, this is what Aharonov seems to conclude, since this statement is reproduced in the 2005 book \cite{Aharonov}. This is also what is discussed in reference \cite{guay}, who asks: {\it{is the Aharonov-Bohm effect due to non locality or  to a long distance effect of fields? I do not see what kind of long distance action of fields could be invoked  except if we admit that Maxwell equations have to be  significantly altered. Non locality, on the other hand is now a well accepted feature of quantum mechanics}}. It is surprising that Aharonov seems to pay no attention to charge conservation; the latter illustrates the statement in reference \cite{bb} quoted above. However, another possibility arises: that the vector potential, the circulation of which on a closed loop leads to a phase factor, has some physical (gauge invariant) reality, although its local value cannot be measured because it is not gauge invariant. The fact that the vector potential becomes a measurable  physical object in a superconducting phase lends some credence to this suggestion, as will be discussed in section (\ref{supercond}).

\subsection{Berry phase, Berry connection, Berry curvature} \label{berry}

In 1984, Berry \cite{berry2} discovered the following: a quantum system in an eigenstate, slowly transported along a circuit $C$ by varying parameters $\bf{R}$ in its Hamiltonian $H(\bf{R})$ acquires a geometrical phase factor $\exp{\left(i\gamma (C) \right)}$ in addition to the familiar dynamical phase factor. He derived an explicit formula for $\gamma(C)$ in terms of the spectrum and eigenstates of $H(\bf{R})$ over a surface spanning $C$. It is a purely geometric object, which does not depend on the adiabatic transport rate around the circuit.

This anholonomy is the quantum analog of the classical one  discussed in section (2.2). Although the system is transported around a closed loop, its final state is different from the initial one. The global phase choice for the initial state is a gauge degree of freedom, which has no effect on Berry's phase. The latter is thus  gauge invariant. A precise definition of adiabaticity is that the motion is slow enough that no finite energy excitation occurs, as is the case when the isolated  system is static. This condition in turn is that a finite excitation gap separates the ground state from the first excited state of the system.

It is conceptually interesting that quantum mechanics establishes a qualitative difference between two sorts of motions:  a motion such that no finite energy excitation occurs, on one hand, and a motion such that finite energy excitations occur. This qualitative difference occurs through a quantitative change of rate of the displacement. Once again, quantity change results in quality change.

One may define the "Berry connection" ${\bf{\vec{\cal A}}}$, the expression of which is given below for completeness:
\begin{equation}\label{connection}
{\cal{A_{\mu}}}({\bf R})\equiv i\langle \Psi_{\bf R}^{(0)}| \frac{\partial}{\partial R_{\mu}} \Psi^{(0)}_{\bf R}\rangle,
\end{equation}
where $|\Psi^{(0)}_{\bf R}\rangle$ is the ground state wave function.  
$\gamma(c)$ is the circulation of ${\bf{\vec{\cal A}}}$ along the curve $C$. ${\bf{\vec{\cal A}}}$ is not gauge invariant since a gradient of any function $f$ can be added to it with no change for $\gamma(C)$. Various generalizations of equation (\ref{connection}) are described in ref.\cite{berry}.

The "Berry curvature" is defined as ${\cal{ \vec{B}}}\equiv \vec{\nabla}\wedge \vec{\cal{A}}$. The Berry phase is equal to the flux of $\cal{ \vec{B}}$ through the closed curve $C$.

There is a close analogy between the electromagnetic vector potential $\vec{A}$ and the Berry connection $\vec{\cal{A}}$.
If a closed box containing a charged particle is driven slowly around a thin solenoid threaded by a magnetic flux, the Berry phase is shown, in this particular case,  to be identical to the Aharonov-Bohm phase.  $\vec{A}$ is shown to be identical to $\vec{\cal{A}}$, modulo the charge coefficient. In fact the derivation of the Berry phase is yet another way of demonstrating the Aharonov-Bohm effect \cite{berry2}.

The ontology of the Berry phase is clear from the numerous experimental observations which have followed various predictions, such as the photon polarization phase shift along a coiled optical fiber \cite{tomita}, Nuclear Magnetic Resonance \cite{suter}, the intrication of charge and spin textures in the Quantum Hall Effects, the quantization of skyrmion charge in a Quantum Hall ferromagnet, etc..\footnote{A more detailed paper on the Quantum Hall Effects is submitted for publication \cite{lederer}}. What remains mysterious is the ontological status of the Berry connection, which is not gauge invariant. The same questions arise about it that are asked about the electromagnetic vector potential.

Whatever the answer to this question, what emerges from the discovery of the Berry phase, and from the many confirmed predictions and  observable effects, connected directly to it\footnote{see for example chapter 4 in ref. \cite{berry}.} etc., is that   doubts about the  reality of the wave function phase as reflecting a profound fact of nature do not appear to be justified.

\section{Superconductivity, a spontaneous breaking of global gauge invariance}\label{supercond}

At low enough temperature, below a "critical temperature" $T_c$,  a number of conducting bodies  exhibit a thermodynamical phase change wherein the resistance becomes vanishingly small, and the diamagnetic response is perfect, which means that an external magnetic field cannot penetrate inside the body (this is called the Meissner effect): a superconducting state is stabilized \cite{tinkham}. This transition is an example of a spontaneous symmetry breaking, whereby the broken symmetry is the global gauge symmetry discussed in the preceding sections. When a continuous broken  symmetry occurs (as is the case of  gauge symmetry in superconductivity), the transition to the new state, which is the appearance of a new quality, is simultaneously characterized by  continuity and  discontinuity. Indeed, the  "order parameter" which describes the new quality grows continuously from zero when the temperature is lowered continuously, but the symmetry is broken discontinuously as soon as the order parameter is non zero. This is easy to understand for a spontaneous breaking of spin rotational symmetry such as that associated with ferromagnetism: as soon as an infinitesimal magnetization appears below the critical temperature, the full rotational symmetry of the microscopic equations, which is the symmetry of the high temperature phase, is broken and rotational symmetry of the state is reduced to an axial rotation symmetry, around the magnetization direction. 


The detailed theory of superconductivity is not relevant here, especially as the microscopic theory of superconductivity in a whole class of new metallic oxydes is still a debated topic. The construction of the  theory of the effect discovered by Kammerlingh Onnes in 1911  lasted half a century. One may surmise that one reason for this delay was precisely the elusive nature of the broken symmetry at work, which was clarified some years after the microscopic theory was published \cite{BCS} in 1957. The superconducting state is characterized by the pairing of a macroscopic number of electrons in so-called "Cooper pairs". A Cooper pair, formed (at least for a whole class of "BCS superconductors"\footnote{In the last thirty years, various other classes of superconductors have been discovered and analyzed, with different ways for electrons to assemble in pairs. This does not limit the generality of the discussion in this paper.}) by two electrons of opposite spins, is a zero spin singlet, and, contrary to electrons, is not a fermion, but a boson. For simplicity, the superconducting ground state can be thought of as the condensation of a macroscopic number of such bosons, where they all have the same phase. Thus the superconducting ground state is characterized by a many-particle macroscopic condensate wave function $\Psi_{SC}({\bf{r}})$, which has amplitude and phase and maintains phase coherence over macroscopic distances.

 How can one reconcile the appearance of a phase\footnote{As remarked by various authors, it is an unfortunate fact that the same word, "`phase"', is used for the "`phase"' of the wave function and for a thermodynamic "`phase"'. I can only hope that this does not produce confusion for the readers of this paper.} in the ground state wave function, which is, as we shall see,  a material object with very concrete and measurable properties --zero resistance and Meissner effect --, with the global gauge symmetry of the many-particle Schr\"{o}dinger equation? This is exactly what spontaneous gauge  symmetry breaking is about: below  the superconducting critical temperature, the state of the system selects a global phase, for the many-particle wave function, which is arbitrary between $0$ and $2\pi$. This is analogous to the ferromagnetic ground state selecting an arbitrary direction in space even though  the Hamiltonian is rotationally invariant. In other words the continuous gauge symmetry of the microscopic equations is spontaneously  broken by the stable thermodynamic state. At the same time  there is  no absolute value for the phase of the wave function of  a single piece of superconductor in free space: any phase can be chosen  between $0$ and $2\pi$.

In the case of ferromagnetism, the direction which the quantum state has picked up for its magnetization can be determined experimentally by coupling it to an infinitesimal test magnetization, such as a compass determining the direction of the earth magnetic field. Similarly, for a superconductor, the phase chosen by the superconducting  ground state can be detected by coupling the superconducting sample to another one: the Josephson effect results (see below); it depends on the phase difference between the two samples.

What about charge conservation in this state? Textbooks state it is violated in the superconducting state. Some philosophers find this hard to believe. It is a matter of understanding orders of magnitude. The explanation is as follows: in order to maintain phase coherence over a macroscopic volume of the superconductor, the total charge fluctuates between macroscopic chunks of the material, by circulation of Cooper pairs, which carry each two electronic charges. The total charge of the sample is conserved, but the charge in a macroscopic part of a sample is determined with an accuracy of  about $10^{-13}$ \cite{tinkham}.  In fact, this is of interest for  whoever still questions the complementarity of canonically conjugate variables (such as $p_x$ and $q_x$ for a single particle with position $x$ and momentum $p_x$). Phase $\phi$ and particle number $N$ are conjugate variables, and the Heisenberg relation holds:
\begin{equation}\label{conjugate}
\Delta N . \Delta \phi \geq 1
\end{equation}
This limits the accuracy with which $N$ and $\phi$ can be simultaneously measured. However, since $N\approx 10^{23}$, and the fluctuation in $N$ is of order $\left(\frac{T_c}{T_F}\right)^{1/2}\sqrt{N}$ an accuracy of $1/\left(\frac{T_c}{T_F}\right)^{1/2}\sqrt{N}\approx 10^{-9}$ on $\phi$ is highly satisfactory and the phase can be viewed as a semi-classical variable. 

The number-phase relationship is expressed in the following relationship\footnote{The detailed expression for the various states involved is not relevant for this paper and can be found, for instance in references \cite{BCS} or \cite{tinkham} for example.}:
\begin{equation}
|\Psi_N\rangle = \int_0^{2\pi} \exp{(-iN\phi/2)}|\Psi_{\phi}\rangle d\phi
\end{equation}
In this equation, $|\Psi_N\rangle, |\Psi_{\phi}\rangle$ are, respectively, the superconducting states for fixed particle number, or fixed phase. The latter is relevant for macroscopic samples. The former is relevant in small superconducting objects, or in the theoretical description of experiments where single  Cooper pairs are manipulated. The factor $1/2$ in the exponential under the summation is due to the fact that Cooper pairs carry 2 electronic charges.

The significance of the phase of the superconducting ground state was not immediately perceived by physicists,  and it took three years to Josephson \cite{josephson}, after the initial BCS paper, to predict that Cooper pairs should be able to tunnel between two superconductors even at zero bias, giving a supercurrent density 
\begin{equation}\label{jc}
J= J_c sin(\phi_1 - \phi_2)
\end{equation}
 where $J_c$ is a constant and $\phi_1, \phi_2$ are the superconducting phases of the two superconductors. Another spectacular prediction was that in the presence of a finite voltage difference between the two superconductors, the phase difference would increase, following equation (\ref{jc}) with time as $2e V_{12}/\hbar$, (where $e$  is the electron  charge and $V_{12}$ the voltage difference) and the current should oscillate with frequency $\omega= 2e V_{12}/\hbar$. As mentionned in ref.\cite{tinkham}, "{\it{Although originally received with some skepticism, these predictions have been extremely thoroughly verified...Josephson junctions have been utilized in extremely sensitive voltmeters and magnetometers, and in making the most accurate measurements of the ratio of  fundamental constants $h/e$\footnote{This was written before the discovery of the Quantum Hall Effects.}. In fact the standard volt is now {\it{defined}} in terms of the frequency of the Josephson effect}}". Among the most well known applications of the effect, SQUIDs (Superconducting QUantum Interference Devices) allow unprecedented accuracy in the detection and measurements of very weak magnetic fields.  Josephson, aged 33,   was awarded the Nobel prize in physics in 1973, for a work done during his PhD.  Subsequently  he worked on telepathy and paranormal phenomena, with no visible success...

\subsection{The London equations}\label{london}

A first significant progress in the theory of superconductivity was due in the nineteen thirties to the London brothers \cite{fhlondon}.
They pointed out that the two characteristic experimental features of superconductivity were zero resistance and the Meissner effect, i.e. the exclusion of magnetic flux by the superconducting body. F. and H. London  proposed two equations which gave a good account of both properties, based on the behaviour of electric and magnetic fields:
\begin{eqnarray}
\bf{\vec{E}}&=& \frac{\partial}{\partial t}\left(\Lambda \bf{\vec{J}}_s \right) \label{1}\\
\bf{\vec{H}}&=& - curl \left( \Lambda \bf{\vec{J}}_s \right)\label{2}
\end{eqnarray}
where $\Lambda$ is a phenomenological parameter found to be equal later to $\frac{m}{n_se^2}$; $m$ is the electron mass, $e$ the electron charge and $n_s $ is the density of superconducting electrons.

Equation \ref{1} describes perfect conductivity since an electric field accelerates  the superconducting electrons, instead of sustaining a constant average velocity, as described by Ohm's law, $\vec{J} = \sigma \vec{E}$ in a normal conductor (where $\sigma$ is the conductivity).

Equation \ref{2} accounts for the Meissner effect, when combined with Maxwell's equation $curl \bf{\vec{H}}= 4\pi \bf{\vec{J}}$. It is straightforward to find that it leads to the exponential screening of the magnetic field from the interior of a superconducting sample, over a distance $\lambda = \sqrt{\Lambda/4\pi}$.

The Meissner effect proves that superconductivity is not equivalent to perfect conductivity, even though equation \ref{1} could be derived in that case. But the magnetic field is not expelled from a perfect conductor.

Both London equations can be condensed in a single one:
\begin{equation} \label{3}
\vec{\bf{J(r)}}_s = -\frac{\vec{\bf{A(r)}}}{\Lambda }
\end{equation}

This equation was actually derived by F. London himself; based on quantum mechanics, but before any microscopic theory, this allowed him to find the expression of $\Lambda= \frac{m}{n_s e^2}$, where $n_s$ is the density of superconducting electrons.

Equations \ref{1} and \ref{2}  can be thought as a good illustration of Duhem's positivism: they account for phenomena, but they say nothing about the ontology of superconductivity. Nothing? In fact, once we admit that magnetic fields are material objects, which they certainly are since they carry energy, the expulsion of field from a bulk superconductor, its exponential disappearance away from the surface within a penetration  length  which is measurable, together with zero electrical resistance, are fundamental real features of the superconducting material. So much so that equations \ref{1}, \ref{2} and\ref{3} have become over the years the hallmarks of any successful microscopic theory of superconducting materials. Their derivation was confirmed after 25 years by a microscopic derivation.

Before discussing equation \ref{3} in more details, it is useful to  mention a more complicated, but more exact expression, due to Pippard, relating the supercurrent and the vector potential.

\subsection{Pippard's equation}\label{pippard}

In normal  conductors, Ohm's law $\vec{J(r)}=\sigma \vec{E(r)}$ can be made  more accurate by noting that the current at a point $r$ is not determined only by the value of the electric field at $r$.  It depends on $\vec{E}(\vec{r'})$ throughout a volume of order $l^3$. $l$ is a length which depends on the scattering processes in a given impure  material.The resulting expression (due to Chambers) is:
\begin{equation}\label{chambers}
\vec{J(\vec{r})} = \frac{3\sigma}{4\pi l}\int\frac{\vec{R}\left[ \vec{R}.\vec{E}(\vec{r'})\right]e^{-R/l}}{R^4}d\vec{r'}
\end{equation}
Where $\vec{R}=\vec{r}-\vec{r'}$. Equation \ref{chambers} reduces to Ohm's law if $\vec{E}(\vec{r'})$ is constant over a distance of order $l$, as is clear by inspection of the formula.


Pippard argued, as was later confirmed from the microscopic theory,  that the superconducting wavefunction should have  a characteritic dimension $\xi_0$ which he found to be $\xi_0 =a\frac{\hbar v_F}{kT_c}$ , where a is a numerical constant of order $1$, $v_F$ is the Fermi velocity and $T_c$ is the superconducting critical temperature.

From the analogy with equation \ref{chambers}, Pippard proposed, in the superconductor,  the replacement of equation \ref{3} by
\begin{equation}\label{pip}
\vec{J_c(\vec{r})} = -\frac{3}{4\pi \xi_0}\int\frac{\vec{R}\left[ \vec{R}.\vec{A}(\vec{r'})\right]e^{-R/\xi}}{R^4}d\vec{r'}
\end{equation}
where $J_c$ is the superconducting current.

The coherence length $\xi$ is related to that of the normal metal  by $\frac{1}{\xi}= \frac{1}{\xi_0} +\frac{1}{l}$. If the normal metal is pure,  $l \approx \infty$.


It is remarkable that the microscopic theory of Bardeen, Cooper and Schrieffer (BCS) \cite{BCS}  justified Pippard's intuition, at the expense of replacing the exponential in equation \ref{pip} by a function $J(R,T)$ which behaves much like the exponential\footnote{For details, see reference \cite{tinkham}.}, and varies smoothly at $R=0$ from $1$ at temperature $T=0$ to $1.33$ at $T=T_c$. 

What matters  here is that the BCS expression reduces to London's equation  \ref{3} if $\vec{A}(\vec{r'})$ is constant over the range of $J(R,T)$, i.e. over a distance of order a few  $\xi_0$.
In other words, Pippard's expression, suitably corrected by the BCS expression differs from London's equation only quantitatively. When $\vec{A}$ varies notably over the range of $J(R,T)$, its  relationship to $\vec{J}(\vec{r})$ is not a simple proportionality, but results from inverting the integral relationship in equation \ref{pip}. 

\subsection{Discussion}
First consider equation \ref{3}.
This equation embodies the essence of superconductivity, and a qualitative change with respect to Ohm's law for a normal conductor.

How can it be physically meaningful if $\vec{A}(\vec{r})$ can change direction, range, magnitude through a gauge change? Obviously a gauge change cannot change the direction, range, magnitude of $\vec{J}$!

 The text book answer to this, for example in reference \cite{tinkham}, p.6, is that London's equation holds only for a particular gauge,  such that the boundary conditions imposed on $\vec{J}$ hold also for $\vec{A}$. For example $div\vec{J} = 0$ is a condition expressing that there is no source from which a superconducting current is created: this is a fact of physics, a fact of the world. So we must also have $div\vec{A}=0$.  Also there is no component of ${\vec{J}}$ perpendicular to the surface of the isolated  superconducting  material. So the implicit answer from text books is that the physical reality of $\vec{J}$ imposes a constraint on $\vec{A}$, the gauge of which is not arbitrary anymore; it is fixed by the physical reality of $\vec{J}$ and is called the London gauge.

Since $\vec{J}$ is a measurable material object, and $\vec{J}$  is proportional to  $\vec{A}$, it seems the latter has transited from the status of (locally) non measurable  object in the normal phase, to that of measurable object of nature in the superconducting phase. Thus, we are led to conclude that equation (\ref{3}) which is a consequence of the global gauge symmetry breaking,  leads to fixing the gauge of the vector potential. The gauge condition on  $\vec{A}$, imposed by the materiality of $\vec{J}$, does not affect the state. The London gauge does not specify the gauge completely, since all harmonic functions $g $ such that $\vec{\nabla}^2g=0$ are possible choices. Here again, the ultimate harmonic  gauge choice is dictated by the superconductor geometry.

We are now facing an interesting ontological question: the vector potential appears to be a material object in a broken global  gauge symmetry phase, such as a superconductor, as evidenced by the proportionality between $\vec{A}$ and the superconducting current density. On the other hand, in the (normal conductor) gauge invariant phase, it has measurable effects only through its circulation on a closed curve. 

How can we get over (aufheben, in german) this contradiction?

Before trying to answer this question, I have to discuss the problem of emergence. This is what the next section is about.

\section{Emergence or reduction? Or both?}\label{emerg}

The topic of emergence in physics, and its antinomy with reductionism, has been discussed  by a number of authors, in particular after Anderson's paper {\it{More is Different}} \cite{PWA2}.
A related  paper by  Laughlin and Pines \cite{pineslaughlin}  has inspired comments by Batterman \cite{baterman} Howard \cite{howard}, Healey \cite{healey2}, among others.

How can the concept of emergence be grounded on rational criteria? Is emergence antinomic with the reductionist approach?

\begin{itemize}
\item Laughlin and Pines (hereafter LP) disagree with the reductionist point of view developed by a number of high energy physicists, such as, for example, Weinberg \cite{weinberg}. The latter author, a particle theorist, defends an ontological reductionism: following him, the fundamental laws that govern elementary constituents of matter ultimately explain phenomena in  all areas of nature. Laughlin and Pines argue that many phenomena of Condensed Matter physics are emergent, and are regulated by what they call "`higher organisation principles"'in nature, which cannot be deduced from microscopics. Experiments, and artful confrontation of theory with experiments (what Hacking calls "`intervening"') is unescapable.  Due to the higher organising principles, of various sorts, emergent phenomena  exhibit insensitivity to microscopics. Examples of such principles are, for example,  renormalisability (critical properties of continuous phase transitions, quantum critical points) or spontaneous symmetry breaking (superconductivity, ferromagnetism, antiferromagnetism, superfluidity). Another higher organisation principle accounts for the stability of the Quantum Hall Effects: the lowest energy excitation of the 2D electron liquid  has an energy gap above the ground state, and electronic localisation   yields a resistivity plateau, the value of which is given by the constant $e^2/h$. The Aharonov Bohm effect yields exactly the measurements of $hc/e$, is also due to a higher organisation principle: the quantum gauge invariance.  LP argue that no approximate treatment from the Schr$\ddot{o}$dinger equation would yield an exact result. And exact treatments are in general impossible.  They refer approvingly  to Anderson's view that "`More is Different"'\cite{PWA2}.

\item Batterman agrees that phase transitions, which he stresses are qualitative changes of state, are emergent phenomena. For him, the mathematical singularities in the thermodynamic potentials are fundamental to point out the qualitative differences between the phases.   I am not fully happy with his paper. He  does not grasp the richness of the notion of "`higher organising principles "`advocated by LP. He specializes in  the Renormalisation Group (hereafter RG) theory and spends some time explaining it\footnote{Batterman insists on the limit $N\rightarrow \infty$. Is this limit  experimentally out of reach, since samples have finite size? In a  sample with dimensions $\approx 1 cm$, the largest correlation length is about $1cm \approx 10^7$  interatomic spaces, which is practically infinite given the experimental accuracy on temperature measurements at a critical point. This is another example of the importance of considering orders of magnitude.}. But he misses some points. For instance, he claims that the RG accounts for the universality of critical exponents. In fact, the RG explains why there are  universality classes, while the molecular field approximation is  universal in giving for the thermodynamic responses  the same exponents near the critical point, independent of the dimensionality of space or of the order parameter for the various materials considered . The mystery has long been the observed irrational values of critical exponents, which disagree with the universal rational exponents of the molecular field theory.

 The liquid-vapour transition as  only example is not  Howard's best choice, since  what the RG explains is why critical exponents depend on two parameters: $n,d$ where $n$ is the dimension of the order parameter of the condensed phase, and $d$ the dimension of the space in which the system is embedded. For the liquid-vapor transition, there is no symmetry breaking: both phases are translation invariant, and the transition with $n=1$ can be discontinuous. When a continuous symmetry is spontaneously broken, a "`higher organisation principle"' such as that which governs the  ferromagnetic ground state entails specific low energy excitations (magnons), as well as universal exponents relevant for $n=3, d=3$ around the critical point. Magnons have no equivalent in the disordered phase.  Batterman misses a crucial point in favor of emergence: at the critical point of a continuous phase transition, a new symmetry appears, which exists only at the critical point, the dilatation symmetry: the system exhibits the same correlations at all length scales. At the critical point, no other length scale in the system, such as interatomic spacing, plays any role. Although the microscopic Hamiltonians had been known in various cases for more than fifty years before the RG theory explained the universality classes of critical phenomena, dilatation symmetry was found only by working out, with the RG,  the correlation functions at the critical point: they decrease algebraically  with distance instead of exponentially away from the critical point. A directly observable consequence of this is the phenomenon of critical opalescence in certain liquid/liquid phase transitions. The critical point paradigm was not deduced  from the microscopic Hamiltonian, but by a procedure both external to the microscopic Hamiltonian, and based on it. This epistemic aspect is the reflection of the ontological emergence. 

\item Howard also addresses the debate of reduction versus emergence. He defines  supervenience, as different from emergence: "`{\it{Supervenience is an ontic relationship between structures. A structure $S_x$, is a set of entities, $E_x$, together with their properties and relations $PR_x$. A structure, $S_B$,  characteristic of one level, $B$, supervenes on a structure, $S_A$,  characteristic of another level $A$, if and only if the entities of $S_B$ are composed out of the entities of $S_A$ and the properties and relations, $PR_B$, of $S_B$ are wholly determined by the properties and relations, $PR_A$, of $S_A$}}"'.
Following Howard, there is no straightforward relationship between reduction and supervenience. For instance, edge states in the (supervenient) fractional Quantum Hall Effects are due to boundary conditions, and do not  allow reduction.

 Emergence can be asserted either as a denial  of intertheoretic reduction or as a denial of supervenience. For example, according to Howard, superfluidity or superconductivity supervene on physical properties at the particle physics level and hence are not  emergent with  respect to particle physics because  supervenience cannot be denied.

Following Howard, entanglement of  quantum  states of two systems\footnote{The wave function $\Psi_{12}$ of two independent systems is factorized as a product of the wave functions of the two systems.}  cannot be reduced to a product of states of the two systems; in general it is a sum of products of eigenstates of both systems.  I  agree with Howard that this is an elementary example of emergent property, which denies reduction. But the reasons for this differ from Howard's argument. Entanglement is  a qualitative change: two entangled fermions, for example, (spin $1/2$ particles) form a boson (spin $0$ or $1$) particles. But one may also argue that entanglement supervenes on quantum particles as a direct consequence of the superposition principle in  quantum mechanics. Entangled particles form a new quantum object, but reappear as particles if the entangled pair is destroyed by some intervention; their properties and relations are "`{\it{  wholly determined by the properties and relations, $PR_A$, of $S_A$}}"` i.e. quantum mechanics and quantum particles.

Supervenience can be viewed as a straightforward materialist statement that all things are formed of material entities which obey the laws of physics. Establishing a logical distinction between emergence and supervenience leaves aside the question of quantity and quality, which is addressed correctly (not in those words)  by Batterman.

Howard  states that the only emergent property in quantum physics is entanglement. He does not discuss LP's arguments, as,  from the definition of supervenience, emergence is eliminated, at least for superconductivity or superfluidity, because the macroscopic wave function\footnote{Superconductivity is a macroscopic quantum phenomenon, not a mesoscopic one.} is built of entangled pairs. From Howard's point of view itself, it should be stressed that the BCS wave function has two levels of entanglement: entanglement of electrons in bosonic singlets, and macroscopic entanglement of states with different singlet numbers: this last feature is essential in establishing phase coherence. The second entanglement type, which expresses the breaking of gauge invariance by the superconducting wave function,  is absent from microscopics. One may say that superconductivity
is doubly emergent: 1) because superposition of states with different particle numbers is absent from gauge invariant states, and 2) it  is a spontaneously broken symmetry phase.

  Howard, contrary to Batterman,  disregards the spontaneous symmetry breaking at work in continuous phase transitions. He disregards, as does Batterman,  the dilatation invariance at the critical point. Superconductivity is a spontaneously  gauge invariance breaking phase. An entangled pair of particles does not break gauge invariance. The disappearance of any resistance, and the Meissner effect, observed  at temperatures below the superconducting critical  point, the Josephson effect, the quantization of flux in vortices, etc., are indeed emergent properties connected with the broken gauge invariance, a change in quality  of the many-particle system due to a change in quantity, as stressed above.
	
Fractional charges of excitations in the Fractional Quantum Hall liquid \cite{laughlin} are qualitatively new entities, which are signatures of emergence. Edge states, responsible for the transport properties, play no role in the classical Hall regime. Emergence can also be due to finite sample boundaries.

\item Healey \cite{healey2} addresses the question of reduction and emergence in Bose-Einstein Condensates (BEC). From this starting point, he raises several interesting questions such as the reduction of classical physics to quantum physics, which I cannot discuss here for want of paper length\footnote{I have suggested elsewhere that classical mechanics, which has no superposition of states, is a case of emergence from quantum mechanics, the theory of which is still to be developed.}. His  questions regarding   the emergence of a phase difference between two different BECs are important ones, etc..  Those questions are specific of BEC physics, perhaps also  in some respects of superconductivity or Josephson effects, but what is interesting  in the context of this paper, he supports the view that spontaneous symmetry breaking is in general a case of emergence.
 
\end{itemize}

Spontaneous  symmetry breaking is a clear example of the category of transformations of quantity in quality \cite{PWA2,pineslaughlin}. Not only because it requires an infinite number of particles but also because it occurs under quantitative changes of parameters such as temperature, pressure, magnetic or electric fields, etc..

In the case of spontaneously broken continuous symmetries (such as ferromagnetism), the Goldstone theorem states that the low symmetry phase possesses collective bosonic modes the energy of which goes continuously to zero with their inverse wavelength. Collective modes, such as spin waves in ferromagnets  are elementary excitations of the broken symmetry phase: they disappear as well defined excitations in the disordered phase. They are well defined  particles in the broken symmetry phase; they   have no equivalent for isolated electrons. 


As far as the reduction/emergence antinomy is concerned, I believe this antinomy has no fundamental root. It is clear that all entities which interact in emergent things, such as spontaneously broken symmetry phases, Quantum Hall states, DNA molecules, central nervous system etc., are built from material entities, the microscopic Hamiltonian of which  is known. In most cases (but not in liquid crystal mesophases for example), they obey quantum mechanics. So reductionism seems vindicated. Not so, because large numbers of interacting things  suffer qualitative changes under suitable external conditions, so that new collective entities with new  causal powers appear, which have no equivalent in the microscopic Hamiltonian, because for instance the symmetries of the latter are lowered  in a broken symmetry phases, or because a gap appears between the ground state and the first excited state when isolated particles have a large ground stae degeneracy (Fractional Quantum Hall State). 

It is legitimate in the early twenty first century to consider that no ab initio calculation, with the most powerful computers available, could deduce the behaviour  of a macroscopic number of interacting  particles from the microscopic Hamiltonian\footnote{This point of view is that of Laughlin and Pines \cite{pineslaughlin}.}.  However, if humanity is granted a sufficient large survival time  over a sufficient number of thousands of years, it is quite possible that totally new technologies  may achieve what we consider to day as impossible. In that sense, I disagree with LP who flirt with notions of unknowable things  and eternal impossibilities in what regards knowledge \cite{pineslaughlin}. This posture  ties them with Kant's transcendental idealism, in contrast with their present decidedly (spontaneously?)  materialistic outlook on physics.

A crucial point needs to be stressed: whoever admits the relevance of the emergence concept has strong reasons to agree that "`more is different"'\cite{PWA2}, i.e. that quantity can transform in quality: this is a radical change from the aristotelian antinomy of those two categories. Dialectical materialism is not far behind...

\section{ London's or Pippard's equation?} \label{lp}

Now let us discuss London's equation \ref{3}.

\begin{itemize}
\item \label{l}
Start with  charge conservation, which is a consequence of global gauge invariance. It  is broken in  superconducting phase, which is a phase with  spontaneously broken gauge  symmetry. As mentioned above  this qualifies somewhat  the statement that gauge invariance in the normal state is a mere "representation surplus". Charge conservation in the gauge invariant  world is  a principle of  matter, analogous to momentum conservation in a translationaly invariant system. A possible way out of the dilemma about the ontological nature of the electromagnetic vector potential would be to view it as an emergent material object in the spontaneously broken global gauge invariant phase, as evidenced by equation \ref{3}.

 The superconducting phase itself is an emergent property (in this case the result of a thermodynamic phase transition) of the many body electron liquid of conductors. The analogy here would be, for instance, with the emergence of ferromagnetism from a macroscopic paramagnetic electronic liquid. But in that case the induction which appears in the ferromagnetic phase is the mere conceptual continuation of the magnetic field in free space. In the gauge invariant phase, however, the vector potential exists only through such anholonomies\footnote{Such as the flux quantization in a vortex.} related to the  Aharonov-Bohm phase. What seems to be in trouble, however, is the  thesis attributing to the vector potential a mere role of mathematical description of  phenomena, with no ontological status. Consideration of equation \ref{l} leads one to suspect that the vector potential $\vec{A}$ has become a bona fide real entity in the superconducting phase.

 Is this discussion about London's equation \ref{3} invalidated by Pippard's equation? 
\item \label{p}
Now turn to this equation \ref{pip}. At first sight, it seems to fatally destroy the conclusions drawn in the paragraph (section \ref{l} above. However, examining Pippard's equation \ref{pip} shows that, just as London's equation, it is not gauge invariant: adding the gradient of any function $f$ to $\vec{A}$ in equation \ref{pip} adds a term to  $\vec{J}$ : $$\vec{J} \rightarrow \vec{J} -\frac{3}{4\pi \xi_0}\int\frac{\vec{R}\left[ \vec{R}.\vec{\nabla}f(\vec{r'})\right]e^{-R/\xi}}{R^4}d\vec{r'} .$$
Pippard's equation \ref{pip}, replaces the direct proportionality of $\vec{A}$ and $\vec{J}$ by a linear relationship. $\vec{J}(\vec{r})$ now depends on a weighted  average on $\vec{A}(\vec{r'})$ over a volume of order $\xi^3$ around $\vec{r}$.

It is straightforward to see that the second term in the equation above does not vanish in general. If, for example $\vec{\nabla}f$ is constant over a range $\xi$, the second "`gauge"' term is obviously non zero. Thus, in order for equation \ref{pip} to hold, one must impose on $\vec{A}$ conditions which are dictated by the inversion of equation \ref{pip}.  The latter yields an expression of $\vec{A}$ as a function of $\vec{J}$ which is not the simple proportionality of equation \ref{3}, but keeps being a linear relationship: multiplying one by some factor  $\lambda$ multiplies the other by the same factor. Boundary conditions imposed by nature and by the sample geometry  on $\vec{J}$ determine  the gauge of $\vec{A}$. It is obvious by inspection of equation \ref{pip} that, depending on how fast $\vec{A}$ varies with $\vec{r'}$ around $\vec{r}$, the direction and magnitude of $\vec{J}(\vec{r})$ will differ from those of $\vec{A}(\vec{r})/\Lambda$, just as the direction and intensity of $\vec{J}$ and $\sigma \vec{E}$ start differing from Ohm's law in equation \ref{chambers} if $\vec{E}$ varies fast enough over a distance $l$. What matters here is that equation \ref{pip} is not a gauge invariant circulation of $\vec{A}$ over a closed loop, but a gauge dependent weighted average on a small volume around $\vec{r}$. In order for equation\ref{pip} to have a meaning, a gauge imposed by conditions on $\vec{J}$ determines constraints on $\vec{A}(\vec{r'})$, and the latter object has to be as real as $\vec{J}(\vec{r})$.

In conclusion, as we have already guessed by noting that equation \ref{pip} may  reduce to London's equation \ref{3} under certain conditions, the complexity introduced by replacing London's equation by Pippard's one does not qualitatively change the conclusions one may draw about the ontology of $\vec{A}$ in the superconducting phase. 
\end{itemize}

  We have seen that real entities such as magnons, which exist in a  ferromagnetic phase do not exist as well defined excitations in the paramagnetic phase. This might suggest a superficial  analogy with the fate of the vector potential, although the latter is not a Goldstone boson. The ontological questions about the magnons in a ferromagnet are of a different category than those about the electromagnetic vector potential in a superconductor, because of the difference between the nature of the broken symmetries in the two cases.  

Further work is needed to understand fully the ontological implications of the fact that the vector potential, which is a sort of ghost in the gauge invariant phase, with only average measurable properties  through its circulation on a closed loop, seems to become a  measurable  entity and definitely appears to reflect the properties of a material object in the broken gauge symmetry phase.

\subsection{The Anderson-Higgs boson}
The recent experimental evidence in favour of the Higgs boson in high energy physics is a major experimental and theoretical achievement. However it came as no big  surprise to physicists in condensed matter physics. Indeed, the massless  Goldstone bosons, also known as "Nambu-Goldstone" particles, which emerge in any continuous broken symmetry phase are not present  in the superconducting phase, because of  electromagnetic interactions. The usual statement is that the Goldstone boson has been absorbed by the massive gauge field  bosonic collective mode \cite{simon}.This was realized by P. W. Anderson and published in {\it{Physical Review }} in April 1963, one year before the papers by Higgs and Brout-Englert \cite{pwa}, although as a qualitative suggestion, with no detailed calculation. The missing ingredients in Anderson's paper were non Abelian fields and   relativity, which do not change qualitatively the mechanism  of the Anderson-Higgs-Brout-Englert boson. 

 This reflects a large degree of conceptual unity of gauge theories. It is fascinating that our present understanding of nature in such different fields  as condensed matter and high energy physics resort to the  same basic theoretical ingredients: gauge theories, spontaneous  symmetry breaking, acquisition of mass by gauge  bosons,  etc..   It suggests also, if this was necessary,  that there is no such thing as a hierarchy of scientific fields of knowledge in physics. The continuous development of knowledge with the related continuous improvement of experimental techniques is as potentially  rich in new discoveries and new physical laws in one field as in another. There are as many new surprises at stake for physicists in improving high energy colliders as there are in atomic physics, condensed matter physics, etc., in reaching lower temperatures, larger magnetic fields, larger pressures, etc..

\section{Conclusions} \label{conclu}

\begin{itemize}
\item If the discussion about the vector potential $\vec{A}$ is limited to normal phases, one may conclude that the potential language -- as opposed to the field language -- is a mere theoretical tool, and gauge symmetry a "description surplus". The spontaneously broken gauge symmetry phases, such as superconductivity, point out the importance of the charge conservation associated to gauge invariance, and suggest that a theory wherein gauge freedom is suppressed, which means   non conservation of charge, lead  to the emergence of $\vec{A}$ as reflecting the properties of  a material object. The understanding and the theoretical description of this object may well  still be  incomplete, but further advances should not invalidate its connection with $\vec{A}$.

\item In section \ref{berry}, I have discussed the Berry phase. Just as the Aharonov-Bohm phase, discussed in section \ref{ab}, the former is yet another entity which supports the thesis that the phase of the wave function is a real object, with experimental testable consequences. Both phases, which depend on the gauge choices,  enter expressions of gauge invariant anhalonomies:  the flux of the Berry curvature, or the flux of a magnetic field through a closed curve.
London's equation \ref{l}, as well as Pippard's \ref{pip} suggest that $\vec{A}$ becomes a real local object, in the gauge imposed by the supercurrent. I am not aware of a thermodynamic phase for which the Berry connection $\vec{\cal{A}}$ would undergo a  transformation  similar to that of $\vec{A}$ in a superconductor, although it seems plausible that a superfluid phase such as those of $^3$He would be a good candidate. 

\item Some concepts have appeared a number of times in this paper. That of emergence is another way of expressing a frequent occurrence in nature: the transformation of quantity in quality. Bind together two spin $1/2$ fermions, they turn into a boson; a large number of microscopic bosons condense in a macroscopic quantum state, the phase of which is  measurable; an adiabatic parallel vector transport over a closed circuit on a sphere results in an anholonomy (Berry phase); an adiabatic parallel transport of a quantum system around a flux tube results in a Aharonov-Bohm phase; cool down a metal, it undergoes  spontaneous broken symmetries of different types, depending on the interactions between the electrons, or on the crystalline symmetry of the atoms: broken translation, rotation, gauge invariance; cool down the universe some three hundred thousand years after the Big Bang, and  a sort of metal insulator transition appears, etc.. The category of emergence encompasses a number of different examples of transformations from quantity to quality. Phase transitions and broken symmetries are one of them. 

As quoted above, ``{\it{More is different}}j'', wrote P. W. Anderson \cite{PWA2} in a brilliant and devastating attack on reductionism\footnote{It is worth quoting the concluding words of this paper: {\it{...Marx said that quantitative differences become qualitative ones, but a dialogue in Paris in the 1920's sums it up even more clearly:

FITZGERALD: The rich are different from us.

HEMINGWAY: Yes, they have more money.}}}. He wrote: {\it{ The ability to reduce everything to simple fundamental laws does not imply the ability to start from those laws and reconstruct the universe. In fact, the more elementary particle physicists tell us about the nature of the fundamental laws, the less relevance they seem to have to the real problems of the rest of science, much less to those of society}}. 

\end{itemize}

{\bf{Acknowledgements.}} I wish to thank my colleagues  of the Laboratoire de Physique des Solides (Universit\'e Paris XI-Campus d'Orsay). Discussions with Jean-No\"el Fuchs, Mark Goerbig, Gilles Montambaux, Fr\'ed\'eric Pi\'echon, Marc Gabay,  Julien Bobroff have been helpful. Opinions and possible misconceptions  expressed in this paper, on the other hand,  are under my only responsibility. I would like to thank Professor Maximilian Kistler  (D\'epartement de Philosophie, IHPST, Universit\'e Paris1), for his attention and stimulating suggestions. Special thanks are due to Jean-No\"el Fuchs for a careful reading of the manuscript and useful remarks.

\end{document}